# Topological plasma transport from a diffusion view*

Zhoufei Liu(刘周费) and Jiping Huang(黄吉平)**

Department of Physics, State Key Laboratory of Surface Physics, and Key Laboratory of Micro and Nano Photonic Structures (MOE), Fudan University, Shanghai 200438, China

*Supported by the National Natural Science Foundation of China under Grants No. 12035004 and No. 12320101004, the Science and Technology Commission of Shanghai Municipality under Grant No. 20JC1414700, and the Innovation Program of Shanghai Municipal Education Commission under Grant No. 2023ZKZD06.
**Corresponding author. Email: jphuang@fudan.edu.cn

Recent studies have identified plasma as a topological material. Yet, these researches often depict plasma as a fluid governed by electromagnetic fields, i.e., a classical wave system. Indeed, plasma transport can be characterized by a unique diffusion process distinguished by its collective behaviors. In this work, we adopt a simplified diffusion-migration method to elucidate the topological plasma transport. Drawing parallels to the thermal conduction-convection system, we introduce a double ring model to investigate the plasma density behaviors in the anti-parity-time reversal (APT) unbroken and broken phases. Subsequently, by augmenting the number of rings, we have established a coupled ring chain structure. This structure serves as a medium for realizing the APT symmetric one-dimensional (1D) reciprocal model, representing the simplest tight-binding model with a trivial topology. To develop a model featuring topological properties, we should modify the APT symmetric 1D reciprocal model from the following two aspects: hopping amplitude and onsite potential. From the hopping amplitude, we incorporate the non-reciprocity to facilitate the non-Hermitian skin effect, an intrinsic non-Hermitian topology. Meanwhile, from the onsite potential, the quasiperiodic modulation has been adopted onto the APT symmetric 1D reciprocal model. This APT symmetric 1D Aubry-André-Harper model is of topological nature. Additionally, we suggest the potential applications for these diffusive plasma topological states. This study establishes a diffusion-based approach to realizing topological states in plasma, potentially inspiring further advancements in plasma physics.

---

Plasma, regarded as the fourth state of matter, comprises unbound ions, electrons, and reactive radicals that exhibit collective behaviors.[1,2] This state can be artificially generated by charging gases with direct or alternating currents, radio-frequency waves or microwave sources. The plasma transport is influenced by particle collisions and modulated by the external electromagnetic field resulting from localized charge concentrations.[3-5] Thus, plasma transport represents a distinctive nonlinear diffusion process.[6-8]

Topological physics is a frontier in the contemporary condensed matter physics.[9,10] It transcends the conventional Landau's symmetry breaking paradigm. Over recent decades, a multitude of topological phases have been both theoretically proposed and experimentally identified.[11-13] Except in condensed matter physics, the classical wave system serves as an ideal platform to implement topological phases, catalyzing the emergence of novel research areas like topological photonics[14-16] and topological acoustics.[17-19] Recently, diffusion system,[20-26] with its dissipation nature, has recently been embraced as a novel platform to reveal topological physics, which has garnered significant interest within the realms of artificial metamaterials and condensed matter physics.[27-33]

In recent years, topological physics has been employed in plasma physics to elucidate certain unique phenomena.[34-37] However, most of these investigations see the plasma as a fluid wherein electromagnetic forces are pivotal. As a result, plasma topology is often interpreted through the magnetohydrodynamics theory, a combination of Navier-Stokes equation and Maxwell's equations. Essentially, plasma is perceived as a classical wave system. To the best of our understanding, the topological plasma transport remains unexplored from a particle diffusion perspective. In this study, we adopt the diffusion-migration equation to characterize the topological plasma transport. Initiating with the double ring model, we successfully establish the anti-parity-time reversal (APT) symmetric one-dimensional (1D) reciprocal model within the coupled ring chain structure. This model is topological trivial so it needs a modification to exhibit topological characteristics. Introducing the non-reciprocity and quasiperiodic onsite potential respectively enable us to construct the 1D Hatano-Nelson model with alternating electric fields and the APT symmetric 1D Aubry-André-Harper (AAH) model, representative of two distinct topological phases. The potential applications for these phases are proposed. Our findings introduce a novel paradigm for interpreting topological states through diffusion, potentially expanding avenues for plasma transport manipulation.

Firstly, it is imperative to derive the governing equation for plasma diffusion. Compared to the conventional diffusion system, the realistic plasma transport manifests greater complexity. This complexity arises from the significant influence of interactions between charged particles and inherent local electromagnetic fields. Generally, the transport of charged particles in plasma can be represented by[1,2]

$$\frac{\partial n}{\partial t} - \nabla \cdot (D\nabla n) \pm \nabla \cdot (\mu \boldsymbol{E} n) + \nabla \cdot (\boldsymbol{v} n) = S \qquad (1)$$

where $n$, $D$, $\mu$, $\boldsymbol{E}$, $\boldsymbol{v}$, and $S$ are the density, diffusivity, mobility, electric field, advective velocity, and external source, respectively. Notably, the sign associated with the third term (migration term) exhibits a positive value for positively charged particles and a negative value for their negative counterparts. For the sake of conciseness, our discussion encompasses only electric fields, overlooking the gas-phase reaction and the advection term. Consequently, plasma transport can be simplified to a diffusion-migration process. By leveraging the Einstein relation, Eq. (1) can be rewritten as

$$\frac{\partial n}{\partial t} - \nabla \cdot (D\nabla n) \pm \nabla \cdot \left[\left(\frac{D\boldsymbol{E}}{T}\right) n\right] = S \qquad (2)$$

where $T$ (in units of V) is assumed to be a constant plasma temperature. We can find that Eq. (2) has the similar form with the conduction-convection equation in heat transfer. So we can borrow the wisdom of some models in thermal diffusion to explore the topological plasma transport.

We now turn our attention to a basic double ring model permeated with plasma,[38] as depicted in Fig. 1(a). Both rings are subjected to a pair of equal-but-opposite circular electric fields, denoted as $\pm \boldsymbol{E}$. The interlayer facilitates particle interchange between the rings. For a clearer visualization, the 3D ring diagram can be transformed into 2D planar channels, with periodic boundary condition imposed at their boundaries (see Fig. 1(b)). Drawing upon Eq. (2), the coupling equations for this double ring model can be expressed as

$$\frac{\partial n_1}{\partial t} = D\nabla^2 n_1 + \left(\frac{D\boldsymbol{E}}{T}\right)\nabla n_1 + h(n_2 - n_1)$$
$$\frac{\partial n_2}{\partial t} = D\nabla^2 n_2 - \left(\frac{D\boldsymbol{E}}{T}\right)\nabla n_2 + h(n_1 - n_2) \qquad (3)$$

where $n_1$ ($n_2$) is the particle concentration of the upper (lower) ring, $D$ is the diffusivity of ring, $h=D_i/(bd)$ is the exchange rate of particles between two rings, and $D_i$ is the diffusivity of interlayer. For simplicity, we focus solely on positive particles. Owing to the ring's periodic structure, we can postulate a plane wave solution for Eq. (3) as $n(x,t)= Ae^{i(kx-\omega t)}$, where $A$ is the amplitude of the density field and $\omega$ is the decay rate. Here, $k=2m\pi/L=m/R$ is the effective wave number, where $m$ is the mode order and can be set as $m=1$. Substituting the plane wave solution into Eq. (3), the effective Hamiltonian for the double ring model can be written as

$$\widehat{H} = \begin{bmatrix} -i(k^2D + h) + \frac{kDE}{T} & ih \\ ih & -i(k^2D + h) - \frac{kDE}{T} \end{bmatrix} \quad (4)$$

Remarkably, this Hamiltonian satisfies the APT symmetry, implying its anti-commutation with the PT operator. As illustrated in Fig. 1(c), the spectrum exhibits an exceptional point (EP)[39] that separates the APT unbroken and broken regions. Within the APT unbroken phase, the simulated density field of the double ring model reaches the steady state quickly. The density profile at the steady state exhibits a minor deviation from the initial one, as shown in Fig. 1(d). Moreover, the phase of the maximum density point reaches a stable state over time, as demonstrated in Fig. 1(e). Conversely, in the APT broken phase, the simulated density field keeps moving and so does the phase of maximum density point, as evidenced in Figs. 1(f) and (g).

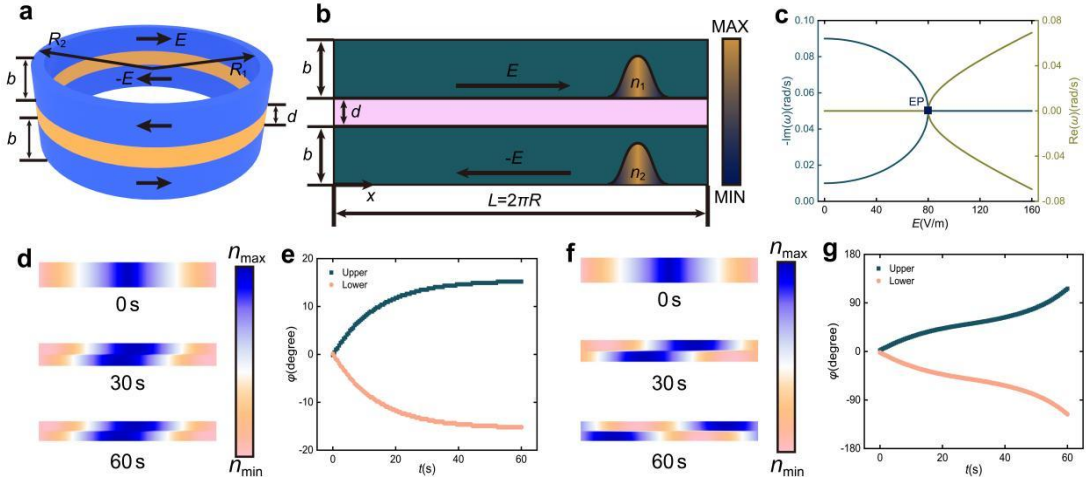

**Fig. 1.** Double ring model. (a) Schematic diagram of double ring model in three dimensions. The rings are indicated in blue while the interlayer is colored in orange. The electric fields in two rings are applied in opposite directions. $b$, $d$, $R_1$, and $R_2$ denote the thickness of ring, the thickness of interlayer, the inner radius of ring, and the outer radius of ring, respectively. Here $R_1 \approx R_2 \approx R$ for approximation. (b) The 2D planar diagram of double ring model. (c) The decay rate and eigenfrequency of the effective Hamiltonian for double ring model. The square indicates the EP. (d) Density field simulation at $E$=40 V/m. Here we choose the uniform excitation as the initial condition: The middle of the channel is set as the highest density $n_h$, while the two boundaries are set as the lowest density $n_l$, and the densities between these two positions are linearly distributed. (e) The phase evolution of the maximum density point for two rings at $E$=40 V/m. Here the middle of channels is chosen as the origin of phase. (f) Density field simulation at $E$=120 V/m. (g) The phase evolution of the maximum density point for two rings at $E$=120 V/m. Parameters: $b$=12.5 mm, $d$=2 mm, $R$=100 mm, $D$=1×10$^{-4}$ m$^2$/s, $D_i$=1×10$^{-6}$ m$^2$/s, $T$=2 V, $n_h$=300 /m$^3$, and $n_l$=200 /m$^3$.

Next, we increase the number of rings to develop the coupled ring chain structure, as depicted in Fig. 2(a). This design aids in the realization of tight-binding model within the diffusion system.[40,41] Additionally, alternating and equal-but-opposite circular electric fields are applied to adjacent rings to maintain the APT symmetry. The equivalent tight-binding model for this structure is shown in Fig. 2(b), representing the APT symmetric 1D reciprocal model. The corresponding Hamiltonian is expressed as

$$\widehat{H} = \sum_{n=1}^{N}\left[-i(k^2D + 2h) + (-1)^n \frac{kDE}{T}\right] a_n^\dagger a_n + ih \sum_{n=1}^{N-1}(a_{n+1}^\dagger a_n + a_n^\dagger a_{n+1}) \quad (5)$$

where $n$ is the site number, $N$ is the number of rings, $a_n^\dagger$ and $a_n$ are the creation and annihilation operators for the lattice model. Except for APT symmetry, this model also satisfies the chiral-time reversal symmetry.[42] The Hamiltonian's imaginary and real spectra are presented in Figs. 2(c) and (d), respectively. As typical in diffusion systems, the only observable decaying branch is the slowest one, with its EP marked by a square in Figs. 2(c) and (d). Consequently, we can only excite the eigenstates on the slowest decaying branch in different phases, as highlighted by the stars in Fig. 2(c). Within the APT unbroken domain, the density field shows a phase-locking effect, exhibiting a lag phase of $\Delta\varphi=\pi/6$ at the steady state, as portrayed in Fig. 2(e). Notably, the field amplitude distribution is uniform during the steady state. Conversely, in the APT broken realm, the density field exhibits an edge localization without a constant phase difference between adjacent channels at the steady state, as illustrated in Fig. 2(f). It is crucial to note that such boundary localization is not topologically protected and can be disturbed by convection disorders. Therefore, in order to achieve the topological model, we should modulate the hopping amplitude and onsite potential respectively onto the APT symmetric 1D reciprocal model.

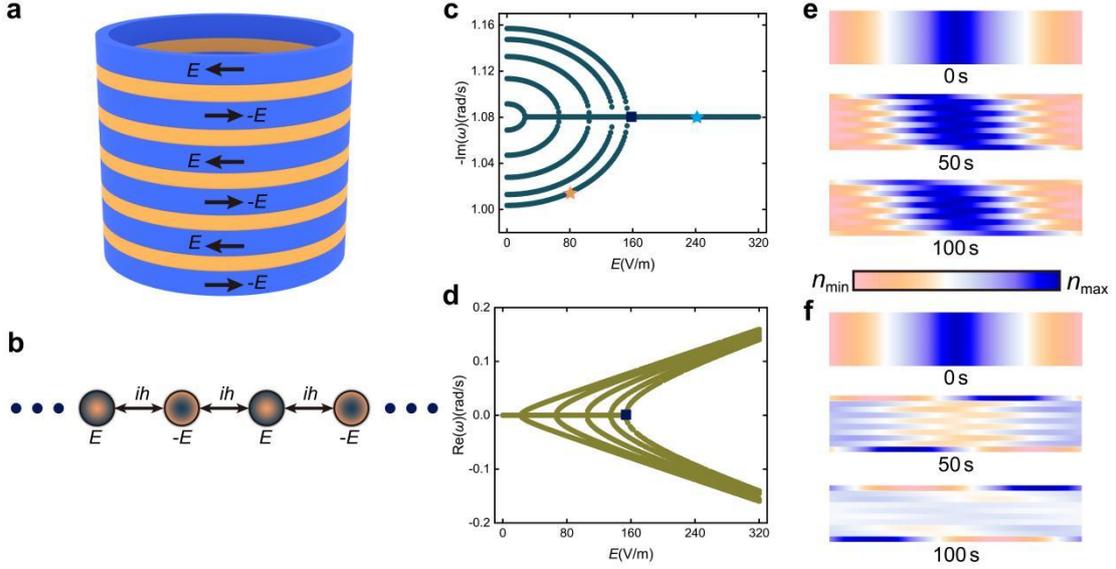

**Fig. 2.** APT symmetric 1D reciprocal model. (a) Schematic diagram of coupled ring chain structure in three dimensions. The clockwise and counter-clockwise electric fields are imposed alternately in adjacent rings. (b) The equivalent tight-binding model. The hopping amplitude between neighbouring sites is $ih$. The (c) decay rate and (d) eigenfrequency of APT symmetric 1D reciprocal model. The square indicates the EP corresponding to the slowest decaying branch. The orange and blue stars denote the slowest decaying branch at $E$=80 V/m and 240 V/m. (e) Density field simulation at $E$=80 V/m with the phase-locking effect. (f) Density field simulation at $E$=240 V/m with a boundary localization and an unlocked phase. Here $N$=10 and other parameters are the same as in Fig. 1.

Firstly, we introduce the non-reciprocal hopping amplitudes into the APT symmetric 1D reciprocal model to realize the non-Hermitian skin effect, which acts as an intrinsic non-Hermitian topology.[43-45] Figure 3(a) shows the equivalent non-reciprocal model, which is the 1D Hatano-Nelson model with alternating electric fields.[46] The effective Hamiltonian is

$$\widehat{H} = \sum_{n=1}^{N} \left[ -iS_0 + (-1)^n \frac{kDE}{T} \right] a_n^\dagger a_n + \sum_{n=1}^{N-1} (iaha_{n+1}^\dagger a_n + iha_n^\dagger a_{n+1}) \qquad (6)$$

where $a$ is an asymmetric factor and $S_0$ is the unified onsite potential of all rings. For this model, we can define a winding number to characterize the eigenvalue topology:[44]

$$W = \frac{1}{2\pi i} \int_0^{2\pi} dk \partial_k \log \det[\widehat{H}(k) - E_B] \qquad (7)$$

where $E_B$ is the base energy, $k$ is the Bloch vector, and $\widehat{H}(k)$ is the Bloch Hamiltonian. For $a$>1 (<1), $W$ is calculated to be $W$=-1 (1). As demonstrated in Figs. 3(b) and (c), the imaginary and real spectra for this non-reciprocal model are similar with the results for APT symmetric 1D reciprocal model in Figs. 2(c) and (d). This correspondence arises because these two models can be transformed with each other via a similar transformation, without affecting the eigenvalues. As illustrated in Fig. 3(d), all eigenmodes for $a$=4 exhibit a localization at the last ring, which is the behaviour of skin effect. In the APT unbroken phase, the density field predominantly accumulates towards a single edge channel, exhibiting a constant phase difference between adjacent channels at the steady state, as showcased in Fig. 3(e). This phenomenon is termed as the phase-locking diffusive skin effect.[47] Conversely, in the APT broken phase, even though the density field retains the diffusive skin effect, the phases of each channel become disordered without any locking, as depicted in Fig. 3(f).

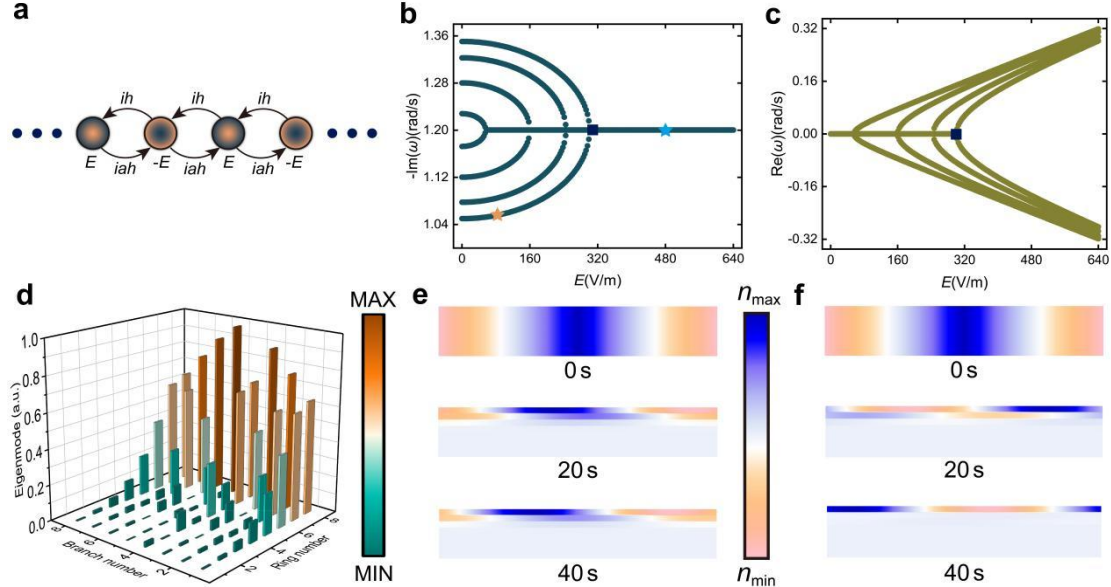

**Fig. 3.** 1D Hatano-Nelson model with alternating electric fields. (a) The equivalent tight-binding model. The hopping amplitudes between neighbouring sites are *ih* and *iah*, where *a* is the asymmetric factor. The (b) decay rate and (c) eigenfrequency of model. The square indicates the EP corresponding to the slowest decaying branch. The orange and blue stars denote the slowest decaying branch at $E$=80 V/m and 480 V/m. (d) Eigenmode distributions of all eigenstates for the non-reciprocal model at $E$=80 V/m. (e) Density field simulation at $E$=80 V/m with the phase-locking skin effect. (f) Density field simulation at $E$=480 V/m with only skin effect. Here $N$=8, $a$=4, and other parameters are the same as in Fig. 1.

In addition to non-reciprocity, we can introduce a complex quasiperiodic onsite potential into the APT symmetric 1D reciprocal model, leading to the realization of a non-Hermitian quasicrystal. This model is the so-called APT symmetric 1D AAH model,[48-50] which possesses inherent topological characteristics.[51,52] The Hamiltonian is represented as

$$\widehat{H} = -i\sum_{n=1}^{N}[V\cos(2\pi\alpha n + i\varphi) + Mh]a_n^\dagger a_n + ih\sum_{n=1}^{N-1}(a_{n+1}^\dagger a_n + a_n^\dagger a_{n+1}) \quad (8)$$

where $V$ is the onsite potential amplitude, $\alpha=(\sqrt{5}-1)/2$ is the inverse golden ratio, and $\varphi$ is the imaginary phase. To accommodate the complex onsite potential, the diffusivities and circular electric fields within the rings require adjustments. Hence, a constant term $Mh$ is added to ensure the adjusted ring diffusivities remain positive. For this particular non-Hermitian quasiperiodic model, an extended-localized transition precisely aligns with the APT transition point $\varphi_c=\log(2h/V)$,[51] as is numerically verified by the decay rates and eigenfrequencies of the model in Figs. 4(b) and (c). In the APT unbroken phase, the eigenstate is extended with an even distribution as shown in Fig. 4(d). While in the APT broken phase, the eigenstate is localized at the certain ring as depicted in Fig. 4(e). Moreover, a topological invariant can be defined to characterize the cooccurrence of APT transition and delocalized-localized transition. By introducing a real phase $\phi$ into the onsite potential, the Hamiltonian can be expressed as

$$\widehat{H}(\phi) = -i\sum_{n=1}^{N}\left[V\cos\left(2\pi\alpha n + \frac{\phi}{N} + i\varphi\right) + Mh\right]a_n^\dagger a_n + ih\sum_{n=1}^{N-1}(a_{n+1}^\dagger a_n + a_n^\dagger a_{n+1}) \quad (9)$$

The winding number is

$$W_\phi = \frac{1}{2\pi i}\int_0^{2\pi} d\phi\, \partial_\phi \log\det[\widehat{H}(\phi) - E_B] \quad (10)$$

$W_\phi = 0\ (\pm 1)$ for the extended (localized) state. By mapping the APT symmetric 1D AAH model onto the ring chain's Hamiltonian, the adjusted parameters of channels are available from the following equations:

$$-i(k^2 D_1 + h) - \frac{kD_1 E_1}{T} = -i[V\cos(2\pi\alpha + i\varphi) + Mh]$$

$$-i(k^2 D_n + 2h) + (-1)^n \frac{kD_n E_n}{T} = -i[V\cos(2\pi\alpha n + i\varphi) + Mh]$$

$$-i(k^2 D_N + h) + (-1)^N \frac{kD_N E_N}{T} = -i[V\cos(2\pi\alpha N + i\varphi) + Mh] \qquad (11)$$

where $n = 2, \cdots, N-1$. For illustration, we get the adjusted diffusivities and electric fields for one extended state ($\varphi$=0.3) and one localized state ($\varphi$=1.2), as shown in Figs. 4(f) and (g). Then we perform the density field simulation for the APT symmetric 1D AAH model. For the extended state in the APT unbroken phase, the profile demonstrates a uniform distribution and remains motionless, as seen in Fig. 4(h). On the other hand, for the localized state within the APT broken phase, the density field distribution reveals multiple moving localization centers, as shown in Fig. 4(i). These simulation results evidently suggest the concurrent emergence of an extended-localized transition and an APT transition in the diffusive APT symmetric AAH model, which is also a topological phase transition.

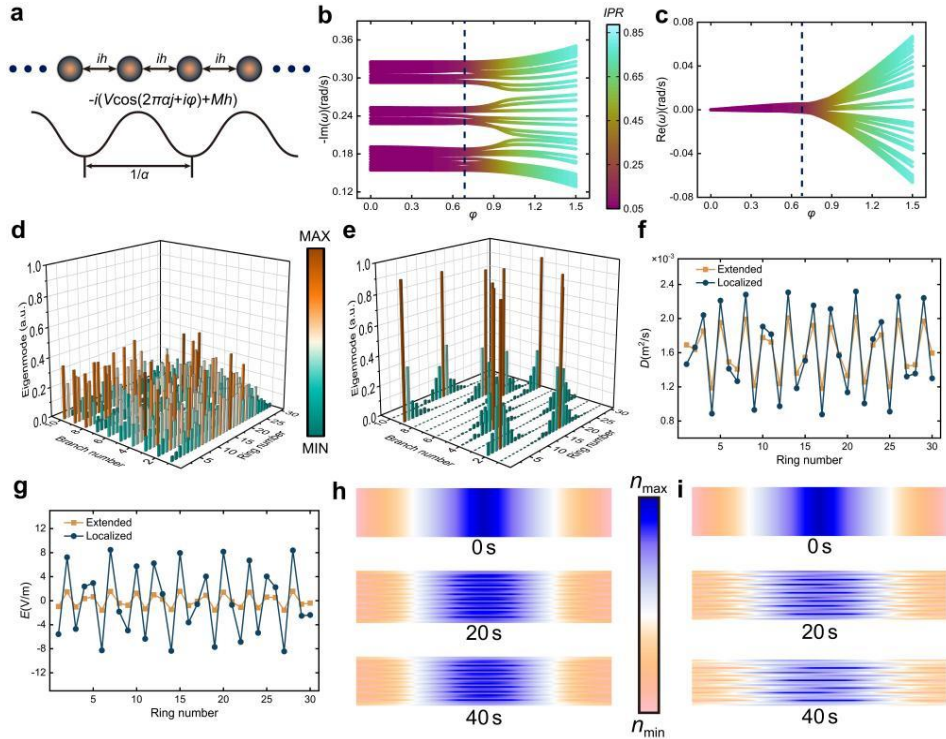

**Fig. 4.** APT symmetric 1D AAH model. (a) The equivalent tight-binding model. The onsite potential is a complex quasiperiodic one. The (b) decay rate and (c) eigenfrequency of model with different $\varphi$. The dashed line indicates the phase transition point $\varphi_c=\log(2h/V)=\log(2)\approx 0.693$. The colorbar means the inverse participation ratio (*IPR*) for each eigenstate. The formula of *IPR* is

$$IPR = \sum_j |\psi_j(E)|^4 / \left(\sum_j |\psi_j(E)|^2\right)^2,$$ where $\psi_j(E)$ is the *j*-th component of the eigenstate

corresponding to energy *E*. *IPR*→0 for an extended state and *IPR*→1 for a localized state. Eigenmode distributions of the ten slowest decaying branches for (d) the extended state ($\varphi$=0.3) and (e) the localized state ($\varphi$=1.2). The adjusted (f) diffusivities and (g) electric fields of each ring for the extended state and the localized state. (h) Density field simulation at $\varphi$=0.3 with a stationary uniform distribution. (i) Density field simulation at $\varphi$=1.2 with multiple moving localization centers. Here *N*=30, $\alpha=(\sqrt{5}-1)/2$, *V*=*h*=$D_i/(bd)$=0.04 /s, *M*=6, and other parameters are the same as in Fig. 1.

Furthermore, the potential applications of diffusive topological plasma transport are promising. The localized position of plasma skin mode can be tailored to specific location via parameter tuning, aiding in flexible plasma manipulation.[30] Besides, the plasma sensor can be achieved because of the high sensitivity of diffusive skin effect to the boundary condition.[53] In catalyst preparation, the plasma localized state, with its high density, can potentially enhance catalytic efficiency. Similarly, this plasma localized state can bolster the efficacy of plasma-assisted engines in aerospace applications. Moreover, the diffusive topological plasma transport offers insights into unique plasma phenomena, paving the way for advanced applications in astrophysics and space physics.[34,35]

In conclusion, our research has utilized the diffusion-migration equation to explore topological plasma transport. We begin with a simple double ring model, illustrating the density field behaviors within APT unbroken and broken phases. Then we increase the number of rings to construct the APT symmetric 1D reciprocal model. However, this model is topologically trivial so we need to adjust the hopping amplitude and onsite potential respectively to introduce topological characteristics. After incorporating the non-reciprocity, we elucidate the density behaviour of non-Hermitian skin effect, which is a non-Hermitian topology. Concurrently, the integration of quasiperiodic onsite potential enables the formation of the APT symmetric 1D AAH model, which has its topological origin. We further put forward the potential applications for the diffusive topological plasma transport. Our diffusion approach offers a potential pathway to uncover more topological states in plasma physics. Besides, this formalism can be compared and complemented with the results based on the magnetohydrodynamics theory.

Nonetheless, it has to be mentioned that the theory of diffusive topological plasma physics needs to be investigated furthermore. To begin with, while deriving the diffusion-migration equation presented in Eq. (2), certain approximations have been adopted. We have overlooked the impact of the magnetic field, advective mechanisms, and gaseous reactions. Furthermore, we assume that the temperature is not time-varied and space-varied. Therefore, addressing these problems within our paradigm remains a direction worthy of consideration. Secondly, manipulating diffusivities and electric fields presents inherent challenges due to the intricate interactions between charged particles and electromagnetic fields. Such challenges might be mitigated with the application of additional theories or methods, such as the particle-in-cell/Monte Carlo collision model[4] or the machine learning approach.[54] In brief, we believe that our work is merely a beginner and increasingly researches on this diffusive paradigm will emerge in the near future.